\documentclass{article}
\usepackage{amsmath}
\usepackage{amsfonts}
\usepackage{amssymb}

\begin{document}

\begin{center}
{\large \textsc{Sudden singularities survive massive quantum particle
production}} \vskip8mm \textbf{John D. Barrow}$^{\,a,}$\footnote{%
E-mail: jdb34@damtp.cam.ac.uk}, \quad \textbf{Ant\^{o}nio B. Batista}$%
^{\,b,} $\footnote{%
E-mail: abrasilb918@gmail}, \quad \textbf{J\'{u}lio C. Fabris}$^{\,b,}$%
\footnote{%
E-mail: fabrisjc@yahoo.com.br}, \\[0pt]
\textbf{Mahouton J.S. Houndjo}$^{\,c,}$\footnote{%
E-mail: sthoundjo@yahoo.fr} \quad \textbf{and} \quad \textbf{Giuseppe Dito}$%
^{\,d,e,}$\footnote{%
E-mail: giuseppe.dito@u-bourgogne.fr} $\,,\,\,$ \vskip8mm $^{\,a}$ DAMTP,
Centre for Mathematical Sciences, University of Cambridge, UK\\[0pt]
$^{\,b}$ Departamento de F\'{\i}sica, Universidade Federal do Esp\'{\i}rito
Santo, ES, Brazil \\[0pt]
$^{\,c}$ Instituto de F\'{\i}sica, Universidade Federal da Bahia, Ba, Brazil 
\\[0pt]
$^{\,d}$ Instituto de Matem\'{a}tica, Universidade Federal da Bahia, Ba,
Brazil \\[0pt]
$^{\,e}$ Institut de Math\'ematiques de Bourgogne, Universit\'e de
Bourgogne, Dijon, France\\[0pt]
\end{center}

\vskip 12mm


\begin{quotation}
\noindent We solve the Klein-Gordon equation for a massive, non-minimally
coupled scalar field, with a conformal coupling, undergoing
cosmological evolution from a radiation-dominated phase to a future sudden
singularity. We show that, after regularisation, the energy of the created
particles is zero and the back-reaction from quantum effects does not change
the evolution of the universe near the future singularity and cannot prevent
the finite-time sudden singularity. \newline
\end{quotation}

\vskip 4mm

\section{Introduction}

The combined data from the anisotropy of the cosmic microwave background
radiation, supernovae type Ia, baryonic acoustic oscillations and matter
power spectrum indicate that the dominant component of the mass-energy
content of the universe must be gravitationally repulsive and driving the
acceleration of the universal expansion. In general, such a component can be
represented by a fluid with an equation of state $p=\omega \rho $, with $%
\omega <-\frac{1}{3}$. If $\omega $ is constant and the spatial sections of
the universe is flat, the recent results of 7-years WMAP observations
indicate that $\omega =-1.10\pm 0.14$\cite{komatsu} at $1\sigma $. This
indicates that the null energy condition is violated, and the expanding
universe may hit a universal singularity in its future evolution, after a
finite proper time. This type of future singularity has been named "big
rip", when there is a divergence in the density of the exotic dark energy
fluid, while the scale factor goes to infinity \cite{caldwell}. This strange
behaviour is connected with the violation of the null energy condition. The
possibility that the null energy condition ($\rho +p\geq 0$) is violated,
and cosmological singularities may occur in the future evolution of the
universe, has been evoked frequently in the literature in the last years. 
The possibility of future singularities was pointed out for the
first time in reference \cite{barrow1} and their occurrence does not require
the violation of the energy conditions. There is also a later discussion in
ref\ \cite{staro}. A recent example of future singularities which does not
violate the null energy condition is the 'big brake' singularity, which
emerges from the DBI action \cite{gorinii,gorini}. The big brake
singularity has the curious property that it can be traversed by a pointlike
particle \cite{gorinif}.

A milder type of finite time singularity is the so-called "sudden
singularity". The sudden future singularity occurs without violation of any
energy condition (so $\rho +p\geq 0$ and $\rho +3p\geq 0$ at all times).
This singularity is characterized by a finite value for the scale factor,
its first time derivative, and for the density, while the second derivative
of the scale factor and the pressure diverge at finite time \cite{barrow1}.
They are singularities of the weak sort discussed by Tipler \cite{tip} and
Krolak \cite{kro}.

In general, it is believed that the fate of the universe near a singularity
(past or future) must be affected by quantum effects arising in the extreme
conditions that exist in its spacetime neighborhood. In the case of the big
rip, this problem has been treated for example, in references \cite%
{brasil1,brasil2,anderson}. In these investigations, it was found that the
quantum effects are important. But, the conclusions concerning the
back-reaction of the quantum effects on the evolution of the universe were
harder to decide unambiguously. In the case of the sudden singularity,
quantum effects were studied in references \cite{barrowbis,houndjobis}, and
the results indicated that quantum effects did not change the evolution near
the singularity. These results were all obtained for a massless scalar field.

It is important to extend these results to the case of massive scalar
fields. Here, we extend earlier studies to the case of a massive scalar
field, non-minimally coupled to gravity, with a conformal coupling
parameter, in a universe which possesses a sudden singularity when quantum
effects are absent. We will show that the singularity is unscathed by the
back reaction from the quantum field. The complexity of the background
evolution of the sudden singularity restricts the possibility of a complete
analytical solution. Hence, we simplify the model by considering two phases:
a primordial radiative phase and a sudden singular phase. Under these
conditions, we can determine the evolution of the quantum fields. In
particular, we can calculate the Bogoliubov coefficients and determine the
energy density of the particles created by quantum effects near the
singularity. After regularisation, we find a null result implying that the
sudden singularity persists, unaffected, by the quantum effects due to the
massive scalar field.

The paper is organized as follows. In the next section, we determine the
master equation for the massive, non-minimally coupled, scalar field. In
section III, we set out the cosmological background containing a sudden
singularity, and in section IV, the master equation is solved. The
Bogoliubov coefficients are determined in section V. In section VI, we
regularise the expression for the energy density, and show that the final
result is zero. In section VII we discuss our conclusions.

\section{The master equation}

We shall investigate quantum creation of massive particles near a sudden
cosmological singularity at finite time. We aim to determine if these
quantum effects are strong enough to change the evolution of the universe
near a classical sudden singularity. In order to do this, we must construct
a specific model. We will concentrate on the evolution of a massive scalar
field, $\phi $, non-minimally coupled to gravity, described by the following
Lagrangian: 
\begin{equation}
\mathcal{L}=\frac{1}{2}\phi _{;\rho }\phi ^{;\rho }-\frac{1}{2}m^{2}\phi
^{2}+\frac{1}{12}R\phi ^{2}.  \label{lagran}
\end{equation}%
The presence of mass takes us one step further than previous work concerning
particle production near sudden singularity \cite{barrowbis}, which
considered only massless scalar particle production. The non-minimal
coupling should strengthen any quantum effects near the singularity, and
also introduces some technical features that facilitate obtaining exact
solutions to the problem.

By variation of the Lagrangian with respect to $\phi ,$we obtain the field
equation : 
\begin{equation}  \label{kg}
\Box \phi +m^{2}\phi -\frac{R}{6}\phi =0.
\end{equation}%
Under variation with respect to the metric, the Lagrangian also gives the
momentum-energy tensor: 
\begin{equation*}
T_{\mu \nu }=\frac{2}{3}\phi _{;\mu }\phi _{;\nu }-\frac{1}{6}g_{\mu \nu
}\phi _{;\rho }\phi ^{;\rho }+\frac{1}{2}m^{2}g_{\mu \nu }\phi ^{2}-\frac{%
\phi }{3}\biggr(\phi _{;\mu ;\nu }-g_{\mu \nu }\Box \phi \biggl)+\frac{1}{6}%
G_{\mu \nu }\phi ^{2},
\end{equation*}%
where 
\begin{equation*}
G_{\mu \nu }=R_{\mu \nu }-\frac{1}{2}g_{\mu \nu }R
\end{equation*}%
is the Einstein tensor.

If we use equation (\ref{kg}) together with the flat Friedmann-Lema\^{\i}%
tre-Robertson-Walker (FLRW) metric for the universe, 
\begin{equation}
ds^{2}=dt^{2}-a^{2}(t)(dx^{2}+dy^{2}+dz^{2}),
\end{equation}%
where $a(t)$ is the expansion scale factor, we find 
\begin{equation*}
\ddot{\phi}+3\frac{\dot{a}}{a}\dot{\phi}+\biggr\{\frac{k^{2}}{a^{2}}+m^{2} +%
\biggr[\frac{\ddot{a}}{a}+\biggr(\frac{\dot{a}}{a}\biggl)^{2}\biggl]\biggl\}%
\phi =0,
\end{equation*}%
where over-dots denote derivatives with respect to the comoving proper time, 
$t$. This equation can be rewritten in terms of the conformal time, $\eta $,
defined by $dt=ad\eta ,$ as 
\begin{equation*}
\phi ^{\prime \prime }+2\frac{a^{\prime }}{a}\phi ^{\prime }+\biggr(%
k^{2}+m\,a^{2}+\frac{a^{\prime \prime }}{a}\biggl)\phi =0.
\end{equation*}%
The primes denote derivatives with respect to the conformal time.

Redefining the scalar field as, 
\begin{equation*}
\phi =\frac{\chi }{a},
\end{equation*}%
we obtain the following equation for $\chi $: 
\begin{equation}
\chi ^{\prime \prime}+(k^2 + m^{2}a^{2})\chi =0.  \label{em}
\end{equation}%
This is the principal equation we will work with from now on.

\section{Two cosmological eras}

The sudden singularity can be described by the following expression for the
expansion scale factor \cite{barrow,barrow2,btsag}: 
\begin{equation}
a(t)=\biggr(\frac{t}{t_{s}}\biggl)^{q}(a_{s}-1)+1-\biggr(1-\frac{t}{t_{s}}%
\biggl)^{n},
\end{equation}%
where $t_{s}$ is the time where the sudden singularity occurs, and $a_{s}$
is the value of the scale factor at this moment. Moreover, $0<q\leq 1$ and $%
1<n<2$ where $q$ and $n$ are free constants and no relation is assumed
between the pressure $p$ and the density $\rho $. We have two asymptotic
phases:

\begin{itemize}
\item Primordial phase, $t \rightarrow 0$: 
\begin{eqnarray}
a &\rightarrow& \biggr(\frac{t}{t_s}\biggl)^q(a_s - 1), \\
\dot a &\rightarrow& \frac{q}{t_s}\biggr(\frac{t}{t_s}\biggl)^{q-1}(a_s - 1),
\\
\ddot a &\rightarrow& \frac{q}{t_s^2}(q - 1)\biggr(\frac{t}{t_s}\biggl)%
^{q-2}(a_s - 1).
\end{eqnarray}

\item Singular phase, $t \rightarrow t_s$: 
\begin{eqnarray}
a &\rightarrow& a_s, \\
\dot a &\rightarrow& \frac{q}{t_s}(a_s - 1), \\
\ddot a &\rightarrow& -\frac{n}{t_s^2}(n - 1)\biggr(1 - \frac{t}{t_s}\biggl)%
^{n-2}.
\end{eqnarray}
\end{itemize}

There is a radiation-dominated primordial phase if $q=1/2$. On the other
hand, in the singular phase, the scale factor and its first derivative
approach constants, and the second derivative, $\ddot{a}$, diverges as $%
t\rightarrow t_{s}$, since $n<2$.

In terms of the conformal time, $d\eta =a^{-1}dt\,$, we have for the scale
factor evolution to leading order:

\begin{itemize}
\item Radiation phase: 
\begin{equation}
a = a_0\eta.
\end{equation}

\item Singular phase: 
\begin{equation}
a = a_s.
\end{equation}
\end{itemize}

The scale factor and its first derivative must be continuous during the
transition of one phase to another. If $a_{s}$ is the scale factor value at
the moment of the transition, and $H_{0}$ the corresponding Hubble
parameter, then the transition moment is given by $\eta
_{t}=1/(H_{0}\,a_{s}) $ and $a_{0}=H_{0}\,a_{s}^{2}$.

The isotropic and homogeneous form we have assumed for the cosmological
evolution of the scale factor, $a(t)$, towards a sudden singularity captures
the essential features of the general solution near such a singularity. On
approach to the sudden singularity as $t\rightarrow t_{s}$, the FLRW
solution has the linear asymptotic form

\begin{equation*}
a\rightarrow a_{s}+q(1-a_{s})(1-\frac{t}{t_{s}}).
\end{equation*}

We can be generalise it to an inhomogeneous metric of the form

\begin{equation}
ds^{2}=dt^{2}-(a_{\alpha \beta }+\tau b_{\alpha \beta }+\tau ^{n}c_{\alpha
\beta }+....)dx^{\alpha }dx^{\beta }  \label{inhomog}
\end{equation}%
on approach to a sudden singularity at $\tau \equiv t-t_{s}=0$, where $%
a_{\alpha \beta },b_{\alpha \beta }$ and $c_{\alpha \beta }$ with $\ \alpha
,\beta =1,2,3$ are functions of the space coordinates and $n$ is a constant
such that $1<n<2$. In the absence of an equation of state, nine components
of the symmetric $a_{\alpha \beta },b_{\alpha \beta }$ and $c_{\alpha \beta
} $ tensors are left independent and arbitrary by the field equations as $%
\tau \rightarrow 0$ and so (\ref{inhomog}) is characteristic of part of the
general solution of the Einstein equations in the vicinity of a sudden
singularity \cite{BCT}. A stability analysis of the FLRW solution has also
been performed by Barrow and Lip \cite{Lip}. For other studies of sudden
singularities of this type, in general relativity and related theories of
gravity, see refs \cite%
{rev,rev1,rev2,sh,dab,noj,odin,cat,dab2,dab3,lake,laz,laz2}.

\section{The solutions for the master equations}

Let us return to the master equation (\ref{em}). First, note that if the
mass is zero, $m=0$, then during both phases the equation becomes, , 
\begin{equation}
\chi ^{\prime \prime}+ k^2\chi =0.
\end{equation}%
The solution is the same during these two phases, and they can be written in
the form of plane waves. Since the solution must be continuous, there is no
final effect in the singular phase, for the trace anomaly or for particle
production.

If the mass is non-zero, the equation during the singular phase takes the
form, 
\begin{equation}
\chi ^{\prime \prime}+ (k^2 + m^{2}a_{s}^{2})\chi =0.
\end{equation}%
The solution is still of plane-wave form, but with a frequency that is
affected by the presence of the mass term. During the radiative phase, the
equation is 
\begin{equation}
\chi ^{\prime \prime}+ (k^2 + m^{2}a_{0}^{2}\eta ^{2})\chi =0.  \label{eq}
\end{equation}%
which can be re-written as 
\begin{equation*}
\ddot{\chi}+\biggr(\omega +\frac{x^{2}}{4}\biggl)\chi=0,
\end{equation*}%
where $x=\sqrt{2ma_{0}}\eta $, $\omega =k^{2}/2ma_{0}$, with dots now
denoting derivatives with respect to $x$. This is a parabolic cylinder
equation, and the solutions can be written as, 
\begin{equation}
\chi =e^{-i\frac{x^{2}}{4}}\biggr[c_{1}\,_{1}F_{1}\biggr(r,s,i\frac{x^{2}}{2}%
\biggl)+c_{2}x\,_{1}F_{1}\biggr(p,q,i\frac{x^{2}}{2}\biggl)\biggl],
\end{equation}%
where the $_{1}F_{1}$ are confluent hypergeometric functions (Kummer
functions), with 
\begin{eqnarray}
r=i\frac{\omega }{2}+\frac{1}{4} &,&\text{ \ }s=\frac{1}{2}, \\
p=i\frac{\omega }{2}+\frac{3}{4} &,&\text{ \ }s=\frac{3}{2},
\end{eqnarray}%
where $c_{1,2}$ are constants.

The confluent hypergeometric functions may be expressed as a series: 
\begin{equation*}
_{1}F_{1}(r,s,z)=1+\frac{r}{s}z+\frac{1}{2}\frac{r(r+1)}{s(s+1)}z^{2}+....
\end{equation*}%
The series represented by the constant $c_{1}$ is even, while the solution
represented by the constant $c_{2}$ is odd.

Let us re-write the solution as, 
\begin{equation}
\chi =c_{1}\chi _{1}+c_{2}\chi _{2}.
\end{equation}%
Considering the series expansion described above we find, 
\begin{eqnarray}
\chi _{1} &=&1-\omega \frac{x^{2}}{2}-\frac{1}{4}\biggr(\frac{1}{12}-\frac{%
\omega ^{2}}{6}\biggl)x^{4}+...\,, \\
\chi _{2} &=&x-\omega \frac{x^{3}}{6}-\frac{1}{4}\biggr(\frac{1}{20}-\frac{%
\omega ^{2}}{30}\biggl)x^{5}+...\,.
\end{eqnarray}%
An important detail is that if we want to take into account the mass in
equation (\ref{eq}) we must consider the series at least until order $x^{5}$
($x^{4}$ for the even series), while if the mass is not taken into account,
we can stop at order $x^{3}$ ($x^{2}$ for the even series). The initial
conditions are imposed for $\eta \rightarrow 0$ ($x\rightarrow 0$) when the
mass is negligible. Considering the expansion for the massless case, and
going back to the notation in terms of the conformal time, we find 
\begin{eqnarray}
\chi &=&c_{1}\chi _{1}+c_{2}\chi _{2}\sim c_{1}\biggr(1-\frac{k^{2}\eta ^{2}%
}{2}\biggl)+c_{2}\frac{\sqrt{2ma_{0}}}{k}\biggr(k\eta -\frac{k^{3}\eta ^{3}}{%
6}\biggl) \\
&\sim &c_{1}\cos k\eta +c_{2}\frac{\sqrt{2ma_{0}}}{k}\sin k\eta .
\end{eqnarray}%
Choosing 
\begin{equation}
c_{1}=\frac{1}{\sqrt{2k}}\quad ,\quad c_{2}=\frac{k}{\sqrt{2ma_{0}}}\frac{i}{%
\sqrt{k}},
\end{equation}%
the solution for $\eta \rightarrow 0$ can be written as, 
\begin{equation}
\chi \sim \frac{1}{\sqrt{2k}}e^{-ik\eta },
\end{equation}%
corresponding to the vacuum initial state.

With those choices for the constants $c_{1}$ and $c_{2}$, we now have the
solution at any time, also for the massive case: 
\begin{equation}
\chi =\frac{1}{\sqrt{2k}}e^{-i\frac{x^{2}}{4}}\biggr[\,_{1}F_{1}\biggr(r,s,i%
\frac{x^{2}}{2}\biggl)+i\sqrt{a}x\,_{1}F_{1}\biggr(p,q,i\frac{x^{2}}{2}%
\biggl)\biggl],  \label{sol-ini}
\end{equation}%
where we have restored the mass in the expressions.

\section{The Bogoliubov coefficients}

Suppose that we approximate the solution during the radiation era by a
massless field. This is equivalent to assuming that we have the massive
solution shown above, but in the limit $\eta \rightarrow 0$. Hence, we have
the following solutions corresponding to the two phases: 
\begin{eqnarray}
\phi _{k}(\eta ) &=&\frac{e^{ik\eta }}{\sqrt{2k}}\quad \quad %
\mbox{(primordial phase)}, \\
\phi _{k}(\eta ) &=&\xi _{01}e^{i\tilde{\omega}\eta }+\xi _{02}e^{-i\tilde{%
\omega}\eta }\quad \quad \mbox{(singular phase)},
\end{eqnarray}%
where $\xi _{01,02}$ are constants and $\tilde{\omega}=\sqrt{%
k^{2}+m^{2}a_{0}^{2}}$.

Now we impose the matching conditions for these two solutions at $\eta =\eta
_{c}$ by requiring continuity of the function and of its first derivative.
We obtain the following relations: 
\begin{eqnarray}
\xi _{01} &=&\frac{1}{2}\frac{1}{\sqrt{2k}}\biggr(1+\frac{k}{\tilde{\omega}}%
\biggl)e^{i(k-\tilde{\omega})\eta _{c}}, \\
\xi _{02} &=&\frac{1}{2}\frac{1}{\sqrt{2k}}\biggr(1-\frac{k}{\tilde{\omega}}%
\biggl)e^{i(k+\tilde{\omega})\eta _{c}}.
\end{eqnarray}%
We notice that when $m=0$, $\xi _{01}=1/\sqrt{2k}$ and $\xi _{02}=0$ and so
the solution is the same in the two phases and there are no particle
production effects.

On imposing the quantisation, we find the following expressions for the two
phases: 
\begin{eqnarray}
\phi _{k}(\eta ) &=&\frac{e^{ik\eta }}{\sqrt{2k}}a+\frac{e^{-ik\eta }}{\sqrt{%
2k}}a^{\dag }\quad \mbox{(primordial phase)}, \\
\phi _{k}(\eta ) &=&(\xi _{01}e^{i\tilde{\omega}\eta }+\xi _{02}e^{-i\tilde{%
\omega}\eta })a  \notag \\
&+&(\xi _{01}^{\ast }e^{-i\tilde{\omega}\eta }+\xi _{02}^{\ast }e^{i\tilde{%
\omega}\eta })a^{\dag }\quad \mbox{(singular phase)},
\end{eqnarray}%
where $a$ and $a^{\dag }$ are the creation and annihilation operators. These
solutions (and their derivatives) are continuous at $\eta =\eta _{c}$. They
can be re-written as, 
\begin{eqnarray}
\phi _{k}(\eta ) &=&\frac{e^{ik\eta }}{\sqrt{2k}}a+\frac{e^{-ik\eta }}{\sqrt{%
2k}}a^{\dag }\quad \mbox{(primordial phase)}, \\
\phi _{k}(\eta ) &=&\frac{e^{i\tilde{\omega}\eta }}{\sqrt{2\tilde{\omega}}}b+%
\frac{e^{-i\tilde{\omega}\eta }}{\sqrt{2\tilde{\omega}}}b^{\dag }\quad %
\mbox{(singular phase)}.
\end{eqnarray}%
Hence, we have 
\begin{equation*}
b=\sqrt{2\tilde{\omega}}(\xi _{01}a+\xi _{02}a^{\dag }).
\end{equation*}%
In this way we find expressions for the Bogoliubov coefficients that connect
the quantum modes during the different phases \cite{birrell}: 
\begin{equation*}
\alpha =\sqrt{2\tilde{\omega}}\xi _{01}\quad ,\quad \beta =\sqrt{2\tilde{%
\omega}}\xi _{02}.
\end{equation*}%
The normalisation condition, 
\begin{equation}
\alpha \alpha ^{\ast }-\beta \beta ^{\ast }=1,
\end{equation}%
is satisfied. When the mass is zero (ie a conformally coupled scalar field),
we have $\alpha =1$ and $\beta =0$. The coefficient $\beta $ is associated
with the created particles \cite{birrell}. Hence, the number of created
particles for each mode $k$ is, 
\begin{equation}
N_{k}=\beta \beta ^{\ast }=\frac{1}{4}\biggr(1-\frac{k}{\tilde{\omega}}%
\biggl)^{2},
\end{equation}%
while the energy of each mode is 
\begin{equation}
\rho _{k}=k\,N_{k}.
\end{equation}%
An integration over all $k$-modes gives, 
\begin{equation}
\rho =\int_{0}^{\infty }\rho _{k}\,d^{3}k=\pi \int_{0}^{\infty }k^{2}\tilde{%
\omega}\biggr(1-\frac{k}{\tilde{\omega}}\biggl)^{2}dk.  \label{rhoef}
\end{equation}%
This expression clearly diverges so it is necessary to regularise it. But,
heuristically, since it is a polynomial expression, it seems clear that
after regularisation we must obtain zero. Hence, the particle production
should not contribute to the energy-momentum tensor and the sudden
singularity is unaffected by these quantum effects

Note that the integral (\ref{rhoef}) admits an analytical solution: 
\begin{eqnarray}
\int\rho_k\,d^3k &=& \pi \int k^2\tilde\omega\biggr(1 - \frac{k}{\tilde\omega%
}\biggl)^2dk = \pi\biggr\{k\sqrt{k^2 + \bar m^2}\biggr(\frac{k^2}{2} - \frac{%
\bar m^2}{4}\biggl) - \frac{k^4}{2}  \notag \\
&+& \frac{\bar m^2}{4}\ln\biggr[2\biggr(k + \sqrt{k^2 + \bar m}\biggl)\biggr]%
\biggl\},
\end{eqnarray}
with $\bar m = m\,a_0$. There is no infrared divergence, but there is a
logarithmic divergence when $k \rightarrow \infty$ (ultraviolet limit).

\section{Regularising the energy}

In order to regularise the expression of the energy, we use the $n$-wave
method expounded in the reference \cite{zeldo}. This method is based on the
Pauli-Villars technique used for quantum field theory in Minkowski
space-time. First, let us write the energy as, 
\begin{equation}
\rho =\int_{0}^{\infty }\rho _{k}(k,m)k^{2}\,dk.
\end{equation}%
Let us define, 
\begin{equation}
^{{}}\rho _{k}^{(n)}\equiv \frac{1}{n}\rho _{k}(nk,nm),
\end{equation}%
where $n$ is a parameter that characterizes the order of the divergence.
From this expression we construct the quantities, 
\begin{equation}
E_{k}^{p}=\lim_{n\rightarrow \infty }\frac{\partial ^{p}\rho _{k}^{(n)}}{%
\partial (n^{-2})^{p}}.
\end{equation}%
The expression for the regularised energy is given by, 
\begin{equation}
\rho _{k}^{reg}=\rho _{k}-E_{k}^{0}-E_{k}^{1}-\frac{1}{2}E_{k}^{2},
\end{equation}%
where $E_{k}^{0}$ eliminates the logarithmic divergence, $E_{k}^{1}$ the
quadratic divergence, and $E_{k}^{2}$ the quartic divergence -- all those
that are normally present in the energy-momentum tensor. This regularisation
of the energy corresponds to a full renormalisation of the coupling
constants, as described in \cite{grib,moste}.

We have, 
\begin{equation}
\rho _{k}=\sqrt{k^{2}+\bar{m}^{2}}-2k+\frac{k^{2}}{\sqrt{k^{2}+\bar{m}^{2}}}.
\end{equation}%
It follows that 
\begin{equation}
\rho _{k}^{(n)}=\rho _{k}.
\end{equation}%
Hence, only the zero-order term survives, and leads to, 
\begin{equation}
\rho _{k}^{ren}=\rho _{k}-E_{k}^{0}=\rho _{k}-\rho _{k}=0.
\end{equation}%
As we suspect, the renormalised energy is zero. There is no effect, and the
quantum phenomena associated with the cosmological dynamics do not change
the character of the sudden singularity or prevent its occurrence.

\section{Conclusions}

In this work we have investigated the fate of the universe near a future
sudden singularity due to quantum particle production effects by a massive
scalar field that is non-minimally coupled to gravity. We have used the
method of calculating Bogoliubov coefficients and the expression obtained
for the energy of the created particles near the singularity is divergent.
This divergence is cured using the standard $n$-wave method and the final
result after regularisation is exactly zero. Hence, we can conclude that the
sudden singularity is robust against quantum effects due to the presence of
a massive scalar field. We have used a simple description for the background
cosmological evolution towards a sudden singularity at finite time that
shares the same time evolution as part of the general solution of the
Einstein equations near such a singularity.

An interesting consequence of this result concerns the trace anomaly. A
massless scalar field conformally coupled to gravity is a particular case of
the problem studied here. For this particular case, the result is the same,
with no quantum effects. Hence, the trace anomaly is absent near the
singularity, as had been speculated in reference \cite{barrowbis}. It must
also be stressed, concerning this massless limit, that the same
regularisation method can be used since there is no infrared divergence even
when the mass is zero.

Others fields should be considered to test the robustness of \ a classical
sudden singularity with respect to quantum effects. However, since the
scalar field and its first derivative are both constant at the sudden
singularity, the Klein-Gordon equation implies that in \ general there will
only be a change in the frequency of the quantum modes as the sudden
singularity is approached. This frequency change alone is unable to create
influential quantum effects. Hence, we can surmise that perhaps the result
found here and in references \cite{barrowbis,houndjobis} can be generalised
further, and they can be applied to other types of fields.

One such a generalisation arises if we consider a non conformal
coupling, with a Klein-Gordon equation given by  
\begin{equation}
\Box \phi +m^{2}\phi -\xi R\phi =0.
\end{equation}%
For a flat FLRW metric, this equation reduces to 
\begin{equation}
\phi ^{\prime \prime }+2\frac{a^{\prime }}{a}\phi ^{\prime }+\biggr\{%
k^{2}+m^{2}a^{2}+6\xi \frac{a^{\prime \prime }}{a}\biggl\}\phi =0.
\end{equation}%
For the radiative phase, the scenario is the same as that studied above,
since $a^{\prime \prime }=0$. For the sudden singularity phase, the
situation is more involved since there is a singularity in $a^{\prime \prime
}$. However, a transformation of the type $\phi =a^{-6\xi }\chi $ (which
reduces to our previous transformation for $\xi =6$ -- the conformal
coupling) can eliminate this singularity, leading to a regular
equation with a damped (anti-damped) harmonic oscillator equation for $\xi >%
\frac{1}{6}$ ($\xi <\frac{1}{6}$). This new term seems to be harmless since
the dissipation (anti-dissipation) effect last\ only for a finite time until
the singularity is reached. Hence, we expect that our previous results hold
even in this more general case.

Another way to consider the problem of quantum avoidance of future
singularities is to solve to Wheeler-de Witt equation in universes
containing future singularities. For big rip and big brake singularities,
this question has been analysed in references \cite{kiefer1,kiefer2}, with
some indications that, at least for the big brake case, the singularity can
be avoided due to quantum effects. For the big rip case the situation is
less clear, see also references \cite{nivaldo,nelson}. It is a natural step
to perform such an analysis for the sudden singularity case, an issue we
hope to consider in the future.

\vspace{0.5cm} \noindent \textbf{Acknowledgements:} We thank CNPq (Brasil)
for partial financial support.

\end{document}